# SIMULTANEOUS SPARK PLASMA SINTERING OF MULTIPLE COMPLEX SHAPES

**Charles Manière [1], Elisa Torresani [1] and Eugene A. Olevsky [1,2],***

[1] Powder Technology Laboratory, San Diego State University, San Diego, USA
[2] NanoEngineering, University of California, San Diego, La Jolla, USA
* Correspondence: eolevsky@sdsu.edu; Tel.: +1- (619)-594-6061 (U.S.A.)

**Abstract:** This work addresses the two great challenges of the spark plasma sintering (SPS) process: the sintering of complex shapes and the simultaneous production of multiple parts. A new controllable interface method is employed to concurrently consolidate two nickel gear shapes by SPS. A graphite deformable sub-mold is specifically designed for the mutual densification of the both complex parts in a unique 40 mm powder deformation space. An energy efficient SPS configuration is developed to allow the sintering of a large-scale powder assembly under electric current lower than 900 A. The stability of the developed process is studied by electro-thermal-mechanical (ETM) simulation. The ETM simulation reveals that homogeneous densification conditions can be attained by inserting an alumina powder at the sample/punches interfaces enabling the energy efficient heating and the thermal confinement of the nickel powder. Finally, the feasibility of the fabrication of the two near net shape gears with a very homogeneous microstructure is demonstrated.

**Keywords:** Spark Plasma Sintering; Complex Shapes; Energy Efficient; Multiple Parts; Multiphysics Simulation; Deformed Interface Method

## Nomenclature

| | |
|---|---|
| $h_i$ | Initial shape height (mm) |
| $h_f$ | Final shape height (mm) |
| $D_i$ | Initial relative density |
| $D_f$ | Final relative density |
| $\vec{J}$ | Electric current density (A/m²) |
| $\vec{E}$ | Electric field (V/m) |
| $J$ | Electric current density norm (A/m²) |
| $E$ | Electric field norm (V/m) |
| $\sigma_{elec}$ | Electric conductivity (S/m) |
| $\rho$ | Density (kg/m³) |
| $C_p$ | Heat capacity (J/(kg.K)) |
| $T$ | Temperature (K) |
| $\kappa$ | Thermal conductivity (W/(m.K)) |
| $\underline{\sigma}$ | Stress tensor (N/m²) |
| $\sigma_{eq}$ | Equivalent stress (N/m²) |
| $\underline{\dot{\varepsilon}}$ | Strain rate tensor (s⁻¹) |
| $\dot{\varepsilon}_{eq}$ | Equivalent strain rate (s⁻¹) |
| $tr(\underline{\dot{\varepsilon}})$ | Trace of the strain rate tensor (s⁻¹) |
| $\varphi$ | Shear modulus |
| $\psi$ | Bulk modulus |
| $Pl$ | Sintering stress (N/m²) |
| $\underline{\mathbb{1}}$ | identity tensor |
| $\theta$ | Porosity |

| | | |
|---|---|---|
| $\dot{\theta}$ | | Porosity rate (s$^{-1}$) |
| $\alpha$ | | Surface energy (J/m²) |
| $r$ | | Grain radius (m) |
| $\dot{e}$ | | First strain rate tensor invariant (s$^{-1}$) |
| $\dot{\gamma}$ | | Second deviatoric strain rate tensor invariant (s$^{-1}$) |
| $n$ | | Power law creep stress exponent |
| $A$ | | Power law creep material deformability (Pa$^{-n}$.s$^{-1}$) |
| $A_0$ | | Power law creep material deformability preexponential constant (Pa$^{-n}$.s$^{-1}$) |
| $R$ | | Gas constant 8.314 J/(mol.K) |
| $Q$ | | Power law creep material deformability activation energy (J/mol) |
| $\varphi_{conv}$ | | Convective heat flux (W/m²) |
| $h_{conv}$ | | Convective heat flux coefficient (W/(m².K)) |
| $T_{water}$ | | Water temperature (300 K) |
| $T_{solid}$ | | Calculated solid temperature (K) |
| $J_{rad}$ | | Surface radiosity (W/m²) |
| $\sigma_s$ | | Stefan Boltzmann constant (5.67E-8 W.m$^{-2}$K$^{-4}$) |
| $\epsilon$ | | Emissivity |
| $G$ | | Thermal irradiation flux (W/m²) |
| $Nr$ | | Refractive index |
| $e_b(T)$ | | Surface radiation produced (W/m²) |
| $refl$ | | Reflected radiative heat flux (W/m²) |
| $\varphi_{rss}$ | | Net inward radiative heat flux (W/m²) |
| $\vec{n}$ | | Normal unit vector |
| $J_n$ | | Imposed normal electric current density (A/m²) |
| $J_c$ | | Contact current density (A/m²) |
| ECR | | Electric surface contact resistance (Ω.m²) |
| $U_i$ | | Contact i face electric potential (V) |
| $\dot{q}_c$ | | Contact heat flux (W/m²) |
| TCR | | Thermal surface contact resistance (m².K/W) |
| $T_i$ | | Contact i face temperature (K) |
| $\vec{u}$ | | Displacement vector (m) |
| $p$ | | External applied pressure (Pa) |
| $\sigma_{elec\ Porous}$ | | Electric conductivity for the porous phase (S/m) |
| $\sigma_{elec\ Dense}$ | | Electric conductivity for the dense phase (S/m) |
| $\kappa_{elec\ Porous}$ | | Thermal conductivity for the porous phase (W/(m.K)) |
| $\kappa_{elec\ Dense}$ | | Thermal conductivity for the dense phase (W/(m.K)) |
| $\rho_{elec\ Porous}$ | | Density for the porous phase (kg/m³) |
| $\rho_{elec\ Dense}$ | | Density for the dense phase (kg/m³) |
| $C_{P\ Porous}$ | | Heat capacity for the porous phase (J/(kg.K)) |
| $C_{P\ Dense}$ | | Heat capacity for the dense phase (J/(kg.K)) |

**1. Introduction**

Spark Plasma Sintering (SPS) has shown a great potential for the sintering of advanced materials thanks to the following characteristics: high pressure (100 MPa with graphite tools), high heating rate (50-1000 K/min), high temperatures (up to 2500 °C) and high pulsed electric current (several thousand amperes) [1–3]. These characteristics are useful conditions for the sintering of nanomaterials. Indeed, the sintering temperature can be lowered through high pressure [4–6], the current effect [7], or the careful control of the sintering final stage [8] (via rate-controlled sintering [9,10], two-step sintering cycle [11], etc.). The use of high electric currents and high heating rates enables the reduction of the sintering time to a less than a minute through the flash (ultra-rapid) spark plasma sintering [12,13] of a very broad range of materials [14] from low melting point metals to ultra-high temperature materials such as silicon carbide [15–17].

Despite these advantages, the fabrication of complex shapes by SPS is challenging. The uniaxial die pressing set up generally restrains the sintered shapes to "2D shapes" or shapes with constant thickness. Shapes with high thickness variations generally experience non-homogeneous densification because the areas of high thickness require more shrinkage than the overall specimen's volume shrinkage [18]. The solution of this problem usually requires multiple punch approaches [19–21] eventually using sacrificial materials to control the relative punches' displacements [22]. However, the efficiency of these techniques is limited when the shapes are very complex. A more general and all shapes inclusive approach called the "deformed interface approach" has been developed [23]. This method is based on an assembly of powders containing a deformable interface which allows the post-sintering separation of complex shape parts and the matching sacrificial shapes. The main powder assembly is a working space with a simple external shape (cylinder, cube, etc.) which can be easily sintered. During sintering, the internal interface follows the displacement of the powders, the initial interface shape is then deformed into the final desired shape. If different powders are employed for the main shape and the sacrificial areas (multi-materials configuration), the deformation of the interface need to be predicted via a finite element code (like in [18]). If the same powder is employed (with the same initial density) for the main shape and the sacrificial areas (mono-materials configuration), the deformation of the interface is easily controllable and can be predicted via the mass conservation-based relation between the initial and final specimen height and relative density:

$$h_i = \frac{h_f D_f}{D_i} \tag{1}$$

In the present work, a new "controllable interface approach" is used for the simultaneous sintering of the two 40 mm nickel gears. This new interface based approach combines the usage of the "deformable interfaces" and adjustable interface electrical/thermal fluxes. This work addresses the challenges of complex net-shaping and stability of the sintering of large size specimens. In addition, to allow the heating and sintering of a 40 mm nickel specimen in a small-scale SPS machine, which is limited in electric current magnitude (1500 A), an energy efficient configuration allowing a very stable heating under low (< 900 A) electric current has been developed. A comprehensive electro-thermal-mechanical simulation of the process has been conducted to study this configuration.

## 2. Theory and calculations

In this section the governing equations and the boundary conditions of the electrothermal-mechanical model used to simulate the energy efficient SPS process are described.

### 2.1 Electro-Thermal-Mechanical model

The ETM simulation of spark plasma sintering is based on a comprehensive model able to predict the Joule heating in the SPS "tooling-specimen column" and the densification of the powder specimen. The Joule heating obeys the following equations of the charge conservation and heat transfer:

$$\nabla \cdot \vec{J} = \nabla \cdot (\sigma_{elec} \vec{E}) = 0 \tag{2}$$

$$\nabla \cdot (-\kappa \nabla T) + \rho C_p \frac{dT}{dt} = \boldsymbol{JE} \tag{3}$$

Where $\boldsymbol{JE}$ is the heat generated by the electric current per unit volume.

The pressure assisted sintering of the powder is simulated via the continuum theory of sintering [24–26]. In this theory, the powder is simulated by a compressible continuum medium. Neglecting the gravity influence and the inertia effects, the local expression of the momentum equation governing the mechanical part of the ETM simulation is:

$$\nabla \cdot \underline{\sigma} = 0 \tag{4}$$

Then, the porous visco-plastic behavior of the material is defined by the stress and strain rate tensors relationship:

$$\underline{\sigma} = \frac{\sigma_{eq}}{\dot{\varepsilon}_{eq}} \left( \varphi \underline{\dot{\varepsilon}} + \left( \psi - \frac{1}{3}\varphi \right) tr(\underline{\dot{\varepsilon}})\underline{\hat{1}} \right) + Pl\underline{\hat{1}} \tag{5}$$

The porosity-dependent expression of the shear $\varphi$ and bulk $\psi$ moduli can be theoretically approximated [24,27]. Additionally, we have developed a method to determine experimentally these moduli by SPS tests [28–30]. For the nickel powder considered in this work, the shear and bulk moduli have been previously experimentally determined [30]; their expressions are:

$$\varphi = \frac{1}{\left( \frac{3}{2} + 5\left( \frac{\theta}{0.55 - \theta} \right)^{1.2} \right)} \tag{6}$$

$$\psi = 0.36 \left( \frac{0.55 - \theta}{\theta} \right)^{0.6} \tag{7}$$

The sintering stress $Pl$ expression (related to capillarity forces) depends on the surface energy $\alpha$, the average particle radius $r$ and on the porosity $\theta$.

$$Pl = \frac{3\alpha}{r}(1 - \theta)^2 \tag{8}$$

In our case, the sintering stress is neglected due to the dominant effect of the high external pressure (30 MPa). The equivalent strain rate in the porous medium is defined as:

$$\dot{\varepsilon}_{eq} = \frac{1}{\sqrt{1-\theta}} \sqrt{\varphi \dot{\gamma}^2 + \psi \dot{e}^2} \tag{9}$$

with the strain rate tensor invariants:

$$\dot{e} = \dot{\varepsilon}_x + \dot{\varepsilon}_y + \dot{\varepsilon}_z \tag{10}$$

$$\dot{\gamma} = \sqrt{2\left( \dot{\varepsilon}_{xy}^2 + \dot{\varepsilon}_{xz}^2 + \dot{\varepsilon}_{yz}^2 \right) + \frac{2}{3}\left( \dot{\varepsilon}_x^2 + \dot{\varepsilon}_y^2 + \dot{\varepsilon}_z^2 \right) - \frac{2}{3}\left( \dot{\varepsilon}_x \dot{\varepsilon}_y + \dot{\varepsilon}_x \dot{\varepsilon}_z + \dot{\varepsilon}_y \dot{\varepsilon}_z \right)} \tag{11}$$

The equivalent stress and strain rate are related to each other by a power law creep equation:

$$\dot{\varepsilon}_{eq} = A\sigma_{eq}^n = A_0 \exp\left( \frac{-Q}{RT} \right) \sigma_{eq}^n \tag{12}$$

with the values for nickel [31,32]: $A_0 = 2.06E - 8 \; MPa^{-n}s^{-1}$, Q = 171.1 kJ mol[-1] and n = 7; and for alumina [18], $A_0 = 8.73E5/T \; MPa^{-n}s^{-1}$, Q = 179.0 kJ mol[-1] and n = 1. The volume change rate and the porosity evolution rate are related to each other via the mass conservation equation:

$$\frac{\dot{\theta}}{(1-\theta)} = \dot{\varepsilon}_x + \dot{\varepsilon}_y + \dot{\varepsilon}_z \tag{13}$$

## 2.2 Boundary and interface conditions

In this section the electro-thermal boundary conditions complemented by the expressions for the electric and thermal contact resistances (ECR and TCR) at the interfaces and the mechanical boundary conditions are described. The external electrode is water cooled, it has been showed that this condition can be modeled by a convective flux [33–35] which obeys the following relationship:

$$\varphi_{conv} = h_{conv}(T_{water} - T_{solid}) \tag{14}$$

We previously [36] determined the convection coefficient $h_{conv}$ to be 200 W/(m²K) for the external electrode boundary.

For the surfaces that radiate towards each other and towards the ambient, the total outgoing radiative heat flux $J_{rad}$ (called radiosity) is defined as the sum of the wall thermal radiation $\epsilon e_b(T)$ and the reflected part (*refl*) of the incoming irradiation $G$ (see fig. 1):

$$J_{rad} = refl + \epsilon e_b(T) = (1 - \epsilon)G + \epsilon(Nr)^2 \sigma_s T^4 \tag{15}$$

The net inward heat flux expression $\varphi_{rss}$ is:

$$\varphi_{rss} = \epsilon(G - e_b(T)) \tag{16}$$

In these "surface to surface radiation" boundary conditions each point radiates in every direction. The graphite emissivity is assumed to be equal 0.8 [36]. The graphite surfaces can be also subjected to the cooling by natural convection flux [37]; this is the case if the SPS test is run under argon. However, we neglect this phenomenon here because the experiments have been carried out under vacuum. The electric boundary conditions are: a ground condition (U = 0) on the lower electrode surface (electric current outlet), an imposed normal current density $J_n$ on the upper electrode surface (electric current inlet) and electrical insulations on all other surfaces $J_n = 0$ :

$$-\vec{n}.\vec{J} = J_n \tag{17}$$

A numerical PID is incorporated [38] to regulate the die temperature via the current density $J_n$ on the upper electrode. The electric and thermal contact resistances (ECR and TCR) in all the internal interfaces of the SPS pressing column obey:

$$J_c = \frac{1}{ECR}(U_1 - U_2) \tag{18}$$

$$\dot{q}_c = \frac{1}{TCR}(T_1 - T_2) \tag{19}$$

Similarly to the conventional SPS configuration [39,40], the ECR and TCR for all the SPS interfaces have a great influence on the distribution of the temperature field [33,41–44]. All ECR and TCR used in this model have been previously determined in Ref. [36].

The mechanical part of the model is utilized for the analysis of the behavior of the powder sample. The bottom and lateral interfaces assume a no penetration displacement condition, the pressure is applied on the upper sample surface. These boundary conditions are:

$$\vec{n}.\vec{u} = 0 \tag{20}$$

$$\vec{n}.\underline{\sigma} = -p \tag{21}$$

The temperature-dependent electro-thermal properties [35,38,41] used in the model are reported in Appendix (Table A). The electrothermal properties of the porous materials are determined through the following relationships based on the effective medium approximation [1,45].

$$\sigma_{elec\ Porous} = \sigma_{elec\ Dense}\left(1 - \frac{3}{2}\theta\right) \tag{22}$$

$$\kappa_{elec\ Porous} = \kappa_{elec\ Dense}\left(1 - \frac{3}{2}\theta\right) \tag{23}$$

$$\rho_{elec\ Porous} = \rho_{elec\ Dense}(1 - \theta) \tag{24}$$

$$C_{P\ Porous} = C_{P\ Dense}(1 - \theta) \tag{25}$$

## 3. Experiment and method

In this section, the experimental conditions are detailed first. Then, the processing of the powder specimen using the "controllable interface approach" is described

### 3.1 Materials and energy efficient SPS conditions

All the experiments were carried out using a Spark Plasma Sintering device (SPSS DR.SINTER Fuji Electronics model 515). The Nickel powder and sintered gear-shape specimens were analyzed by the scanning electron microscopy (FEI Quanta 450, USA). The microstructures of the gears where analyzed on polished and etched (50 ml HNO$_3$; 50 ml acetic acid; 50 ml H$_2$O) surfaces at 5 kV (to detect the surface details). Two powders are considered, a nickel powder for the fabrication of the gears (Cerac, Ni 99.9 % pure, 5 μm, see SEM images in Figure 1) and an alumina powder (Materion, $Al_2O_3$ 99.2% pure, 325 mesh) to electrically insulate the nickel powder in the energy efficient configuration.

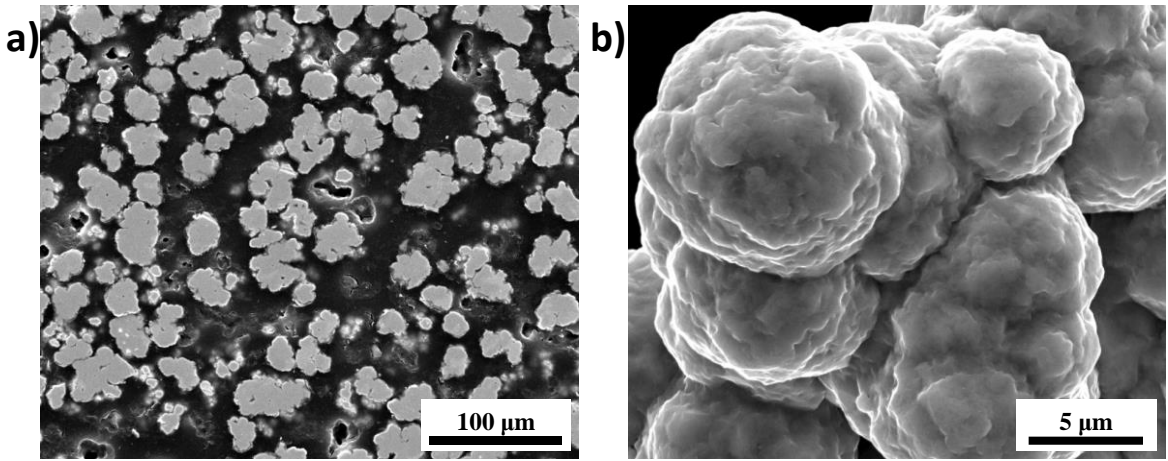

Figure 1: SEM images of as received nickel powder with a) polished and b) grains surfaces.

The SPS column stack is represented in Figure 2. The 40 mm diameter metallic sample (60 mm outer die diameter) cannot be heated in traditional SPS configuration as our device is limited to 1500 A. The principle of the energy efficient configuration is to channel the electric current in a very narrow graphite foil area, which dissipates high energy [46]. This constrained electric current area becomes the main heating element within the utilized configuration and allows for a significantly lower electric current the heating of a large volume of graphite and powders. To generate this special electric current distribution, a 2 mm thick alumina powder is inserted between the two nickel/punch contacts to deviate the electric current in the double layer graphite foil (0.4 mm thick) at the punch/die and sample/die interface. To constrain the electric current in the graphite foil, the die is electrically insulated by a boron nitride sprayed at the foil/die interface. The SPS experiment is carried out at 30 MPa (from the start) and for a heating rate of 40 K/min followed by 5 K/min in the final stage of the sintering. A K type thermocouple was used to measure and regulate the die temperature at the location indicated in Figure 2 (the measurement is not conducted at the die's mid-height to avoid breaking the die). The outcome of this configuration is discussed in the results section.

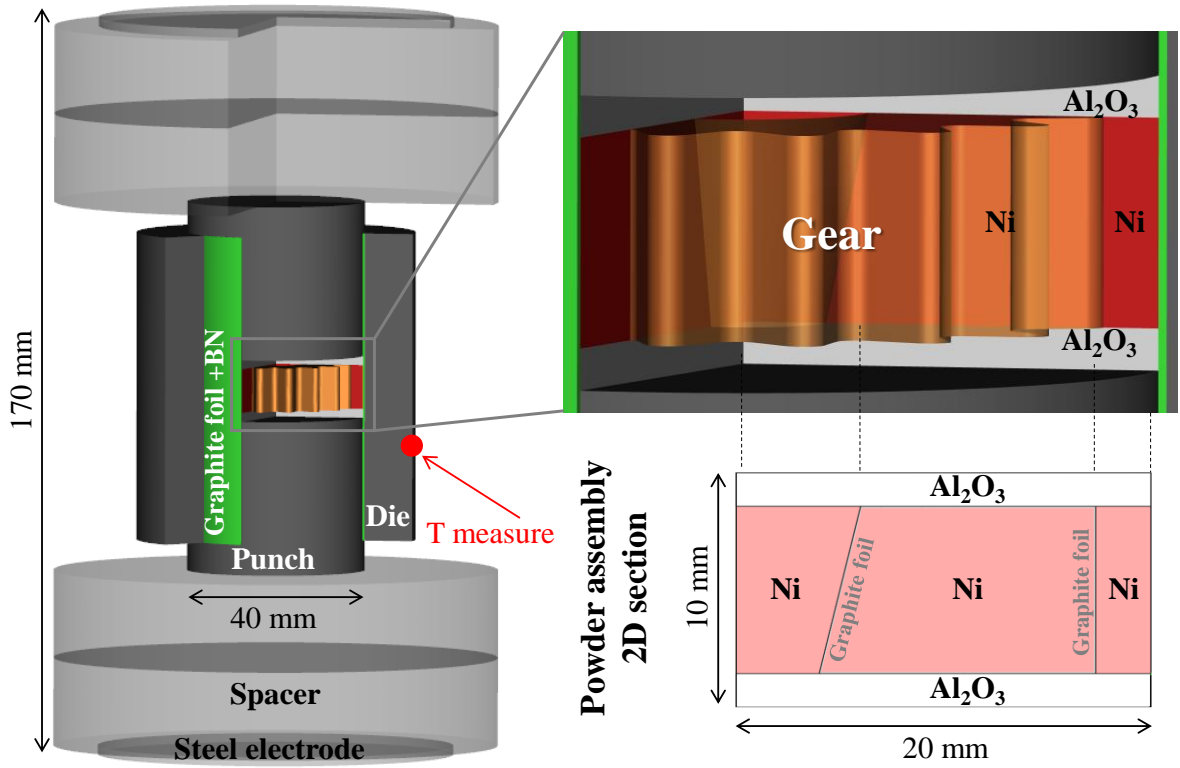

Figure 2: Energy efficient spark plasma sintering configuration with: the overall graphite tooling column; and inside the powder area, the details of the "controllable interface method" for creating nickel gear shape with graphite foil and alumina interfaces.

*3.2 The deformed interface method*

The two gear complex shapes can be obtained simultaneously via a graphite deformable sub-mold which helps the post-sintering separation of the adjacent parts [23]. The immersed graphite foil location in the nickel powder is represented in Figure 2. The concept of this graphite deformable sub-mold approach is simple. The semi-rigid and compressible foil follows the densification of the nickel powder up to the full densification of the overall nickel specimen. After that, the separation of the two gear-shape parts and the central conic hole is possible due to the low mechanical resistance of the graphite foil as compared to the SPS-sintered nickel. The only difficulty of this approach is to impose the gear shape onto the graphite foils and maintain this shape during the loading of the nickel powder. To assist this difficult step, we use a 3D printed polymer gear shape (GEEETECH prusa i3 with ABS filament) to help imposing the gear shape upon the graphite foil and at the same time facilitate the loading step. The different steps are the following: 1) the polymer gear part is inserted in the die, 2) the graphite foil is then imposed onto the gear surfaces; then, 3) the nickel powder is filled first in the central hole and the outer gear area; finally, 4) the polymer part is slowly removed and replaced by the nickel powder. At the end of these stages the nickel powder assembly with an internal complex shape foil interface is generated. The alumina powder is located at the upper and lower sample/punch interfaces and is introduced after these steps. The energy efficient SPS configuration, which allows the sintering of complex shapes, is thereby assembled.

## 4. Results

The stability of the energy efficient configuration is discussed first with the help of the electro-thermal-mechanical simulation; the discussion on the experimental outcome is provided afterwards.

*4.1 Electro-Thermal-Mechanical simulation of the energy efficient SPS process*

The obtained die temperature, displacement, applied electric current and pressure curves are reported in Figure 3. During the 40 K/min ramping, the sintering starts at about 350 °C and starts to decrease at 700 °C (to end at 850 °C in 5 K/min). The range of the sintering temperatures is then substantially extended for this powder. The 40 mm metallic sample has been successfully sintered under applied electric currents lower than 900 A while in the traditional SPS approach these diameters always require several thousand amperes. Thus, we have the confirmation that the developed process is energy efficient and allows the full densification of the 40 mm metal specimen in a lab-scale SPS device (which is limited to 1500 A). The electro-thermal-mechanical (ETM) simulation is then employed to analyze this experiment.

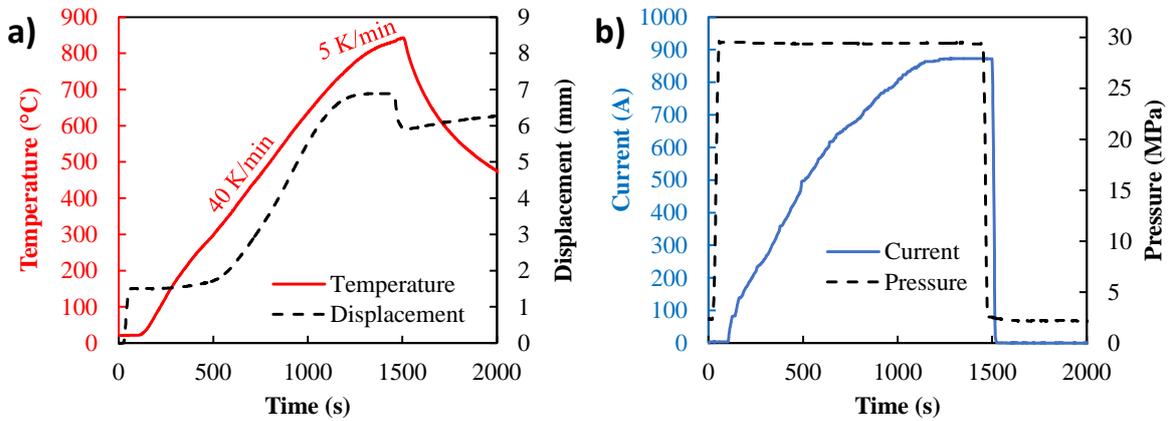

Figure 3: Obtained experimental data from the spark plasma sintering experiment with a) temperature and displacement, b) pressure and electric current.

The compared temperature/current simulated versus experimental curves (reported in Figure 4) showed a relatively good accordance provided that all the input data come from literature. The only important difference is between the experimental and simulated electric current curves in the time interval from 100 to 500 sec. In this region, the electric current (PID regulated) is higher than the experimental current which apparently means that the electric contact resistance ECR in the range of the low temperatures is probably slightly underestimated. However, the error is acceptable because in this time range the sintering of the powder is not yet started. In figure 4 the simulated volumetric loss density and the electric current lines are represented too. The electric current lines show that the area in the vicinity of the contact alumina/die is the area of electric current constriction and it is also the area of the maximum heat dissipation. These two electric current constriction zones can be considered to be the main heating elements within this configuration. In order to determine whether these external heating elements provide a sufficiently stable heating at 40 K/min for this configuration, the heating profile is analyzed. The temperature and relative density fields' evolution during sintering are presented in Figure 5 (see simulation video in *supplemental materials "S1"*). The first analysis of the powder relative density indicates that the temperatures do not reach the values allowing a significant densification of the alumina which is only consolidated in the area in contact with nickel. This allows an easy separation of the gear and the alumina which is a sacrificial element. The simulated temperature field provides the two main observations:

the thermal contact resistances (TCR) and the specific electric current path makes the die and the nickel powder to be the preferentially heated areas;

the heat in the nickel powder area is very homogeneous (ΔTmax ~ 25 K) and seems to start from the edge.

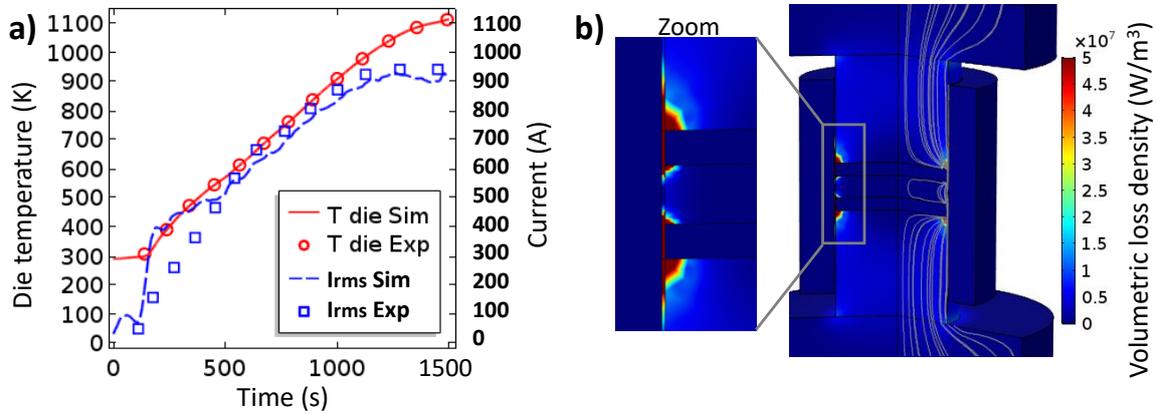

Figure 4: Compared experimental/simulated die temperature and electric current (a), simulated electric current lines and volumetric loss density (b).

The homogeneity of the nickel sample can be explained by the thermal confinement from the lateral TCR and the upper and lower alumina powder which has a very low thermal conductivity (5.23 *vs* 27.5 W.m$^{-1}$.K$^{-1}$ for Al$_2$O$_3$ *vs* Ni at 25°C and 0.938 *vs* 24.6 W.m$^{-1}$.K$^{-1}$ for Al$_2$O$_3$ *vs* Ni at 900°C). The higher conductivity of the metal powder also contributes to the homogenization of the specimen heating coming from the edge. The association of the TCR, the low and high thermal conductivity of the metal *versus* alumina powders makes this energy efficient configuration to be very stable thermally. Therefore the relative density field is in turn very homogeneous during the whole processing cycle.

The controllability of the interface is an important object of future research.

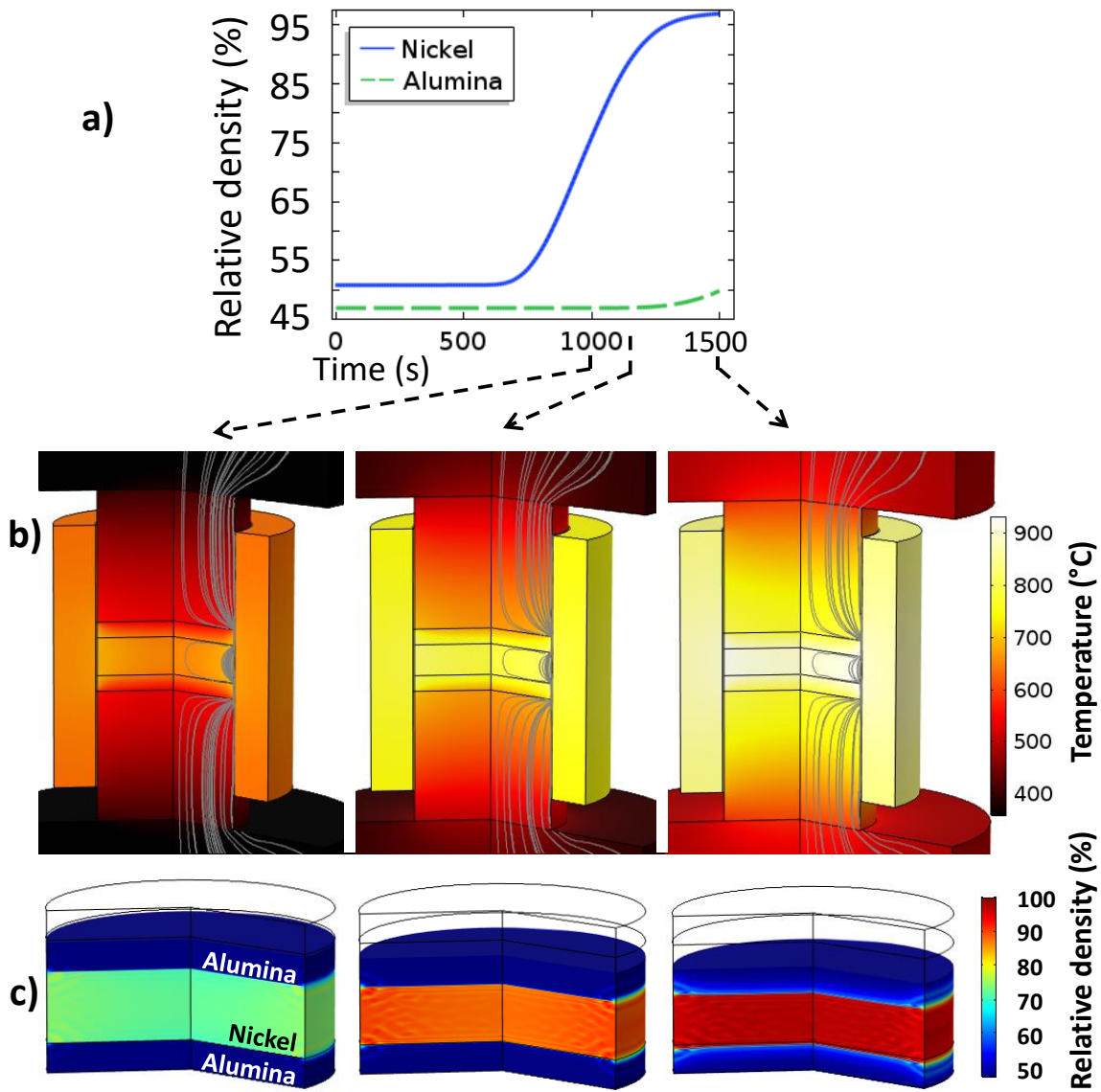

Figure 5: Electrothermal-mechanical simulation results, with in the upper part (a), the mean densifications curves for the alumina and nickel powders and, in the lower part, the simulated temperatures, electric current lines (b) and sample relative density fields (c) at 1000, 1150 and 1500 sec.

*4.2 Fabricated parts and microstructures*

Capture images of the overall process are reported in Figure 6. One can clearly see that the alumina powder in contact with the nickel specimen has little consolidation (as predicted by the simulation in Figure 5) and almost no mechanical strength, which allows its easy removal. The separation of the inner and outer gears was very easy due to the presence of the graphite foils which helps the two parts to slide with respect to each other. After polishing of the surfaces in contact with the alumina powder, the two gears are close to the near net shape quality. The straight angles and round shapes are well reproduced. One can still notice some minor distortions on some gear teeth. This is due to a slight moving of the graphite foil during the powder loading. This can be improved using a more sophisticated supporting tool for the graphite foil interface shaping and the loading of the powder. The surface roughness is of the similar magnitude as the roughness of the powder interface or of the foil in contact. The polished and etched microstructures in the inner gear (center of the specimen) and outer gear (edge of the specimen) are reported in Figure 7. In accordance with the very homogeneous temperatures predicted by the simulation (Figure 5), the microstructure is the same from the center to the edge of the sintered nickel specimen. The grain size (~5 µm) is very similar to the initial nickel particle size suggesting no noticeable grain growth occurring. The presence of a certain level of remaining residual porosity is clearly observable. The shapes of the initial powder particles (Figure 1), which are highly agglomerated, can generate non ideal powder packings [28,47] which may result in these kinds of porous structures.

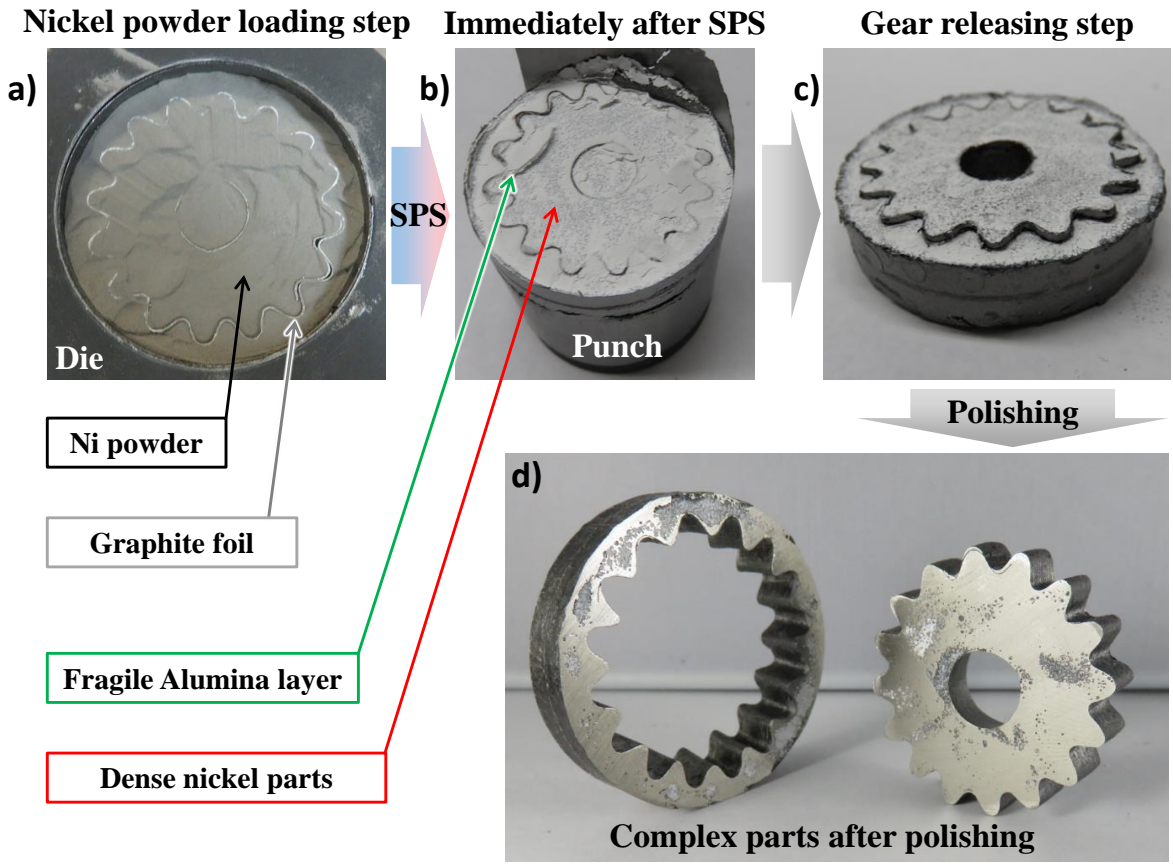

Figure 6: Photos of the main steps of the spark plasma sintering experiment, from the powder loading to the obtained complex shapes; a) loading, b) after SPS, c) releasing and d) the parts after polishing.

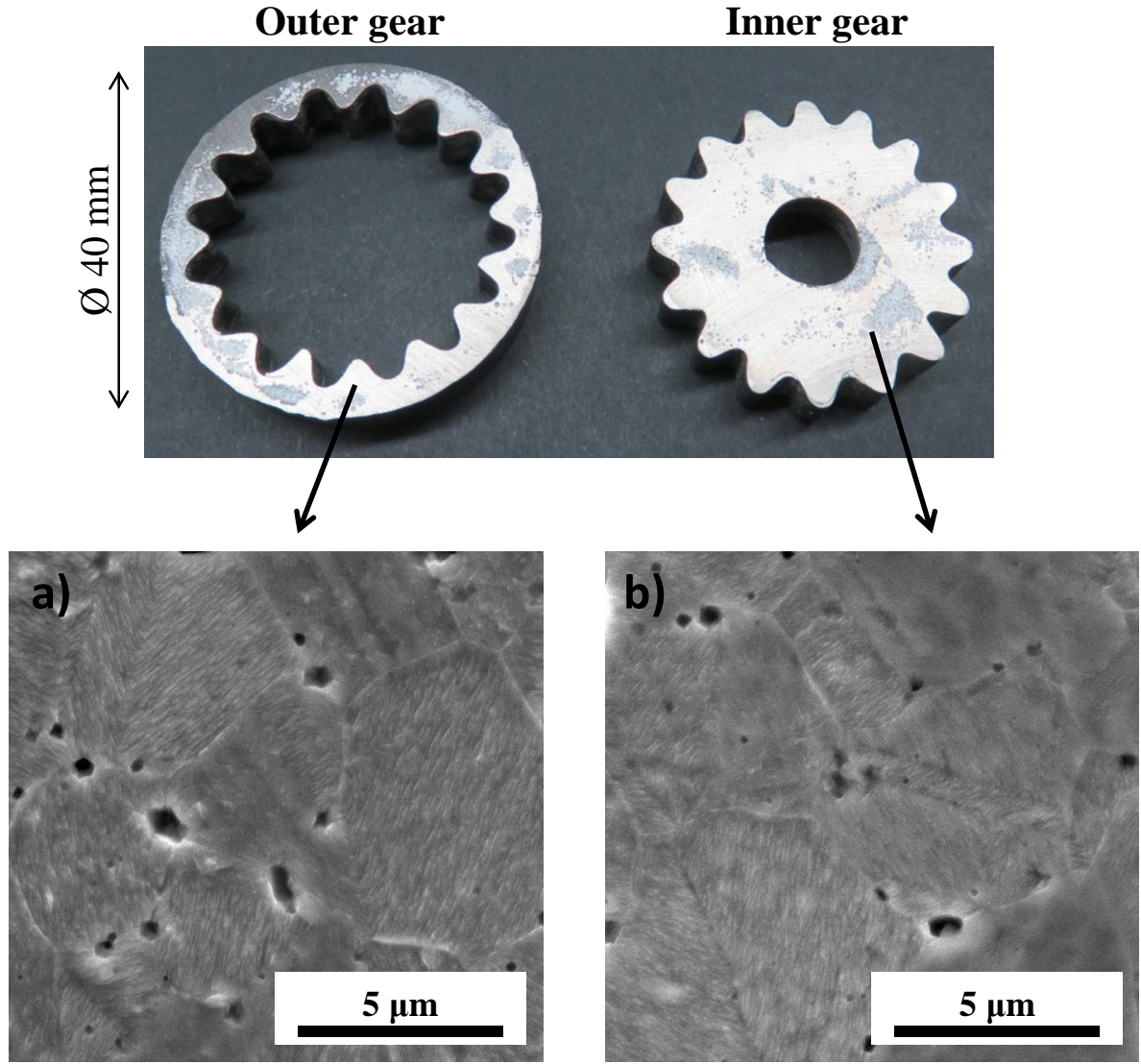

Figure 7: Polished and etched SEM microstructure for the (a) outer gear (edge of the SPS specimen) and (b) the inner gear (center of the SPS specimen).

## 5. Conclusions

An advanced spark plasma sintering approach enabling the fabrication of large size complex shapes and an energy efficient processing is developed. A unique configuration using deformable and electric insulated interfaces has been employed to constrain the electric current path for a significant reduction of the electric current required to heat the 40 mm specimen while imposing a gear shape onto the processed parts. A comprehensive electrothermal-mechanical simulation of the SPS process has been conducted. This simulation reveals that the high thermal stability observed for this SPS approach originates from different factors: the high thermal conductivity of the metal powder and the thermal confinement of this metal powder via the lateral thermal contact resistance and via the upper and lower alumina powder which has a very low thermal conductivity.

Finally, two nickel gears have been obtained with a homogeneous microstructure and an accuracy of the shapes closed to the "near net shape" quality. This case study demonstrates the high potential of the SPS technology which can cumulate the benefits of advanced material properties (through high pressures, high heating rate) and complex shapes. The use of the "controllable/deformable interface approach" modifies the traditional utilization of the SPS technique where one die/punch tooling set is dedicated to the production of one sample. It is theoretically possible to place any number of complex shape interfaces in a large dimensions tooling set (like in a conventional furnace) improving drastically the productivity of the SPS technology. Such an improvement would require a perfect control of the temperatures and a homogeneous displacement field of the powder. The present work is a first step in this direction.

**Supplementary Materials:** The following are available online at www.mdpi.com/xxx/s1, Video S1.

**Author Contributions:** Conceptualization, C.M.; Methodology, C.M., E.T., E.O.; Validation, E.O., E.T.; Formal Analysis, C.M., E.T., E.O.; Investigation, C.M., E.T., E.O.; Supervision, E.O.

**Funding:** The support of the US Department of Energy, Materials Sciences Division, under Award No. DE-SC0008581 is gratefully acknowledged.

**Conflicts of Interest:** The authors declare no conflict of interest. The funders had no role in the design of the study; in the collection, analyses, or interpretation of data; in the writing of the manuscript, or in the decision to publish the results.

**Appendix A**

Table A: Electrothermal materials properties.

| Materials | | Expressions |
|---|---|---|
| Graphite | | $34.3+2.72 \cdot T-9.6E-4 \cdot T^2$ |
| Electrode | $Cp$ | $446.5+0.162 \cdot T$ |
| Alumina | (J·kg$^{-1}$·K$^{-1}$) | 850 |
| Nickel | | $1.01E+02 \cdot \log(T)-1.43E+02$ |
| Graphite | | $123-6.99E-2 \cdot T+1.55E-5 \cdot T^2$ |
| Electrode | $\kappa$ | $9.99+0.0175 \cdot T$ |
| Alumina | (W·m$^{-1}$·K$^{-1}$) | $39500\ T^{-1.26}$ |
| Nickel | | $-8.06E-08\ T^3+2.50E-04\ T^2-2.32E-01\ T+1.34E+02$ |
| Graphite | | $1904-0.0141 \cdot T$ |
| Electrode | $\rho$ | 7900 |
| Alumina | (kg·m$^{-3}$) | 3899 |
| Nickel | | 3965 |
| Graphite | | $1/[1.70E-5-1.87E-8 \cdot T+1.26E-11 \cdot T^2-2.46E-15 \cdot T^3]$ |
| Electrode | $\sigma_{elec}$ | $1/[(50.2+0.0838 \cdot T-1.76E-5 \cdot T^2) \cdot 1E-8]$ |
| Alumina | (S/m) | $1/[8.7E+19\ T^{-4.82}]$ |
| Nickel | | $-4.53E-02\ T^3+1.27E+02\ T^2-1.18E+05\ T+3.85E+07$ |